\documentclass{article}

\usepackage{amsfonts,amssymb,epsfig,epsf,amsthm}
\usepackage{amsmath}
\usepackage{graphicx}

\begin{document}

\newcommand{\BM}{Bohmian mechanics} %
\newcommand{\wf}{wave function} %

\newcommand{\RRR}{\mathbb{R}} %
\newcommand{\CCC}{\mathbb{C}}
\newcommand{\SSS}{\mathbb{S}} %
\newcommand{\EEE}{\mathbb{E}}
\newcommand{\NNN}{\mathbb{N}}
\newcommand{\bo}{\mathrm{B}}
\newcommand{\el}{\mathrm{F}} 
\newcommand{\pos}{\mathrm{pos}} 
\newcommand{\inter}{\mathrm{int}}       
\newcommand{\new}{\mathrm{new}} 
\newcommand{\hi}{${\cal H}$}
\newcommand{\LL}{{\cal L}}
\newcommand{\Fock}{{\cal F}} 
\newcommand{\perm}{\pi} 
\newcommand{\anotherperm}{\varrho} 
\newcommand{\fiop}{\hat{\varphi}}       
\newcommand{\dom}{{\cal D}}     
\newcommand{\1}{\boldsymbol{1}} 
\newcommand{\sign}{\,\mathrm{sign}\,} 
\renewcommand{\div}{\,\mathrm{div}\,} 
\renewcommand{\leadsto}{\to} 

\newcommand{\tx}{\widetilde{x}}
\newcommand{\ty}{\widetilde{y}}
\newcommand{\tw}{\widetilde{w}}
\newcommand{\tX}{\widetilde{X}}
\newcommand{\tY}{\widetilde{Y}}
\newcommand{\vD}{\vec{D}} %
\newcommand{\vA}{\vec{A}}%
\newcommand{\vx}{\boldsymbol{x}}
 \def\Anti{{\rm Anti\,}}
\def\Im{{\rm Im\,}}
\def\Re{{\rm Re\,}}
\def\Sym{{\rm Sym\,}}

\def\D{{\rm d}}
\def\E{{\rm e}}
\def\I{{\rm i}}

\newcommand{\bx}{\boldsymbol x}
\newcommand{\bxp}{\boldsymbol x'}
\newcommand{\bkp}{\boldsymbol k'}
\newcommand{\bg}{\boldsymbol f}
\newcommand{\bgx}{\boldsymbol X}
\newcommand{\ba}{\boldsymbol a}
\newcommand{\bk}{\boldsymbol k}
\newcommand{\bko}{k_0}
\newcommand{\by}{\boldsymbol y}
\newcommand{\be}{\boldsymbol e}
\newcommand{\bb}{\boldsymbol b}
\newcommand{\bc}{\boldsmbol c}
\newcommand{\gre}{\frac{e^{ik|\bx-\bxp|}}{|\bx-\bxp|}}
\newcommand{\bby}{\boldsymbol Y}
\newcommand{\bbx}{\boldsymbol X}
\newcommand{\ab}{|A|}
\newcommand{\dtb}{|\Delta T|}
\newcommand{\dt}{\Delta T}
\newcommand{\bom}{\omega}

\pagenumbering{arabic}

\title{On the exit statistics theorem of many particle quantum scattering}

\author{Detlef D\"urr\\Mathematisches Institut der Universit\"at M\"unchen, Germany\\duerr@mathematik.uni-muenchen.de\\[3mm]
Stefan Teufel\\Zentrum Mathematik, TU M\"unchen\\teufel@ma.tum.de}

\date{July 22, 2003}

\maketitle

\begin{abstract}
We review the foundations of the scattering formalism for one
particle potential scattering and discuss the generalization to
the simplest case of many non interacting particles. We point out
that the "straight path motion" of the particles, which is
achieved in the scattering regime, is at the heart of the crossing
statistics of surfaces, which should be thought of as detector
surfaces. We sketch a proof of  the relevant version of the many
particle flux across surfaces theorem and discuss what needs to be
proven for the foundations of scattering theory in this context.

\end{abstract}

\section{Introduction}
Quantum mechanical scattering theory is usually
about the $S$-matrix. The operator $S$ maps the so called
in-states $\alpha$  to  out-states $\beta$. That may seem
sufficiently self explanatory as basic principle since

\begin{quote} \it An experimentalist generally prepares a state \ldots
at $t\to -\infty$ and then measures what this state looks like at
$t\to +\infty$. \\[1mm]    \rm {\footnotesize S.\ Weinberg in  ``The quantum theory
of fields'' \cite{W}, Chapter 3.2: The S-Matrix}
\end{quote}
and
\begin{quote} \it The $S$-matrix $S_{\alpha,\beta}$ is the probability
amplitude for the transition $\alpha \rightarrow \beta$ \ldots \\[1mm]\rm
{\cite{W} \footnotesize Chapter 3.4: Rates and Cross Sections. }
\end{quote}
so everything seems settled. However the quote continues

\begin{quote}\it
\ldots but what does this have to do with the transition rates and
cross sections measured by experimentalists? \ldots

\ldots we will give a quick and easy derivation of the main results,
actually more a mnemonic than a derivation, with the excuse that
(as far as I know) no interesting open problems in physics hinge
on getting the fine points right regarding these matters\ldots \\[1mm] \rm
{\footnotesize Chapter 3.4: Rates and Cross Sections.  }
\end{quote}

The mnemonic recalls that the plane waves in the $S$-matrix
formalism are limits of wave packets but it does not come to grips
with the time dependent justification of the scattering formalism,
in fact it does not connect to the empirical cross section.

We remark aside, that apart from not making contact with the
empirical cross section,  there is another---though quite
related---problem with the mnemonic, which---as is felt by
many---can only be settled by interesting new physics: When a
particle is scattered by a potential its wave will be spread all
over. What accounts then for the fact, that a point particle event
is registered at one and only of the detectors? Where did the
particle come from which is suddenly manifest in that detector
event? This is some facette of the measurement problem of orthodox
quantum theory \cite{Bell,Against}. We shall not say more on that
in this paper and refer to \cite{DGTZphysica}. We emphasize
however that we shall use Bohmian mechanics for a theoretical
description of the cross section---a theory free from the conceptual problems
of quantum mechanics.

We immediately jump now to the technical heart of foundations of
scattering theory by observing that
$$t\to \pm \infty$$
means the {\bf mathematical} limit of the formulas capturing the
{\bf physical} situation (see (\ref{lim}) below). Experimentalists
prepare and measure states at {\bf large} but {\bf finite}  times.
 They count the
number of particles entering the detectors.  The physical meaning
of the $S$-matrix derives from being  the limit expression of the
theoretical formula for the number count.  It is moreover
immediately clear---once this point of the finiteness of the
physical situation has been recognized---that the times at which
particles cross the detector surfaces are random. The detector
clicks when the particle arrives.  That time is random and not
fixed by the experimenters. Thus the foundations of quantum
mechanical scattering theory become slippery: No observables
exist, neither for time measurements nor for position measurements
at random times. The question is thus: What are the formulas which
theoretically describe the empirical cross section and which
result in the appropriate limit in the $S-$matrix formalism?

In this paper we shall shortly review the simple one particle
potential scattering situation. Apart from discussing the quantum
flux we shall introduce Bohmian mechanics, which allows to capture
the theoretical foundations of scattering theory in the most
straightforward way.  We shall then extend our considerations to
 multi particle potential scattering and show, why
the multi-time flux (which we shall introduce) determines the
statistics in this case in terms of a generalized  flux across
surfaces theorem. The first paper  on the flux across surfaces
theorem \cite{CNS} discusses also the multi particle flux but
restricts the computation of statistics to the marginal statistics
of one particle only, ignoring thus the most important
correlations due to entangled wave functions. Our multi-time
analysis deals specifically with entangled wave functions.

\section{The theoretical cross section }

We   adopt  conventional units in which
$\frac{\hbar}{m}=1$ and recall that the theoretical prediction
$\sigma_{\mathbf{k}_0}(\Sigma)$ for the cross section as given by
$S$-matrix  theory is
\begin{equation}\label{thpre}
 \sigma_{k_0}(\Sigma)=
16\pi^4\hspace{-0.8pt}\int\limits_{\Sigma}\D\omega \,
  |T(|k_0|\omega,k_0)|^2\,.
  \end{equation}
Here $T=S-I,$ where the identity $I$
   subtracts the unscattered particles from the scattered
 beam. As to be explained below, (\ref{thpre}) is based on a model for a beam of particles.
 Using heuristic stationary methods,
  Max Born \cite{Born} computed $T$ in the first paper on quantum
mechanical scattering theory. We shall
recall his argument shortly, since it serves on its
own as defining a theoretical cross section.

 Consider solutions $\psi$ of the stationary
Schr\"odinger equation with the asymptotics
\begin{equation}\label{naive}
\psi(x) \approx \E^{\I k_0\cdot x} +
f^{k_0}(\omega)\frac{\E^{\I|k_0||x|}}{|x|}\,\quad {\rm
for} \,|x|\,{\rm large}
\end{equation}
and $x=\omega|x|$.
In naive scattering
theory  the first term is regarded as representing an incoming
plane wave and the second term as the outgoing scattered wave with
angle-dependent amplitude.

Such wave functions  can be obtained as solutions of the
Lippmann-Schwinger equation \begin{equation} \label{LS} \psi(x,k)
= \E^{\I k\cdot x} - \frac{1}{2\pi}\int\D y\, \frac{ \E^{\I|k||x-y|}}{|x-y|} \,V(y)\,\psi(y,k)\,.
\end{equation} The  solutions form a complete set, in the sense
that an expansion in terms of these generalized eigenfunctions, a
so-called generalized Fourier transformation, diagonalizes the
continuous spectral part of $H$.  Hence the $T$-matrix can be
expressed in terms of generalized eigenfunctions and one finds
(cf.\ \cite{RS3}) that
\begin{equation}\label{TV}
T(k,k') =
(2\pi)^{-3} \int\D x\, \E^{-\I k\cdot x}\, V(x) \,\psi(x,k')\,.
\end{equation}
Thus the iterative solution of (\ref{LS}) yields  a perturbative
expansion for $T$, called the Born series.

Moreover, comparing (\ref{naive}) and (\ref{LS}), expanding the
right hand side of (\ref{LS}) in powers of $|x|^{-1}$, we see from
the leading term  that
\[
f^{k_0}(\omega) = -(2\pi)^{-1}\int\D y\, \E^{-\I|k_0|\omega\cdot y}\, V(y)\, \psi(y,k_0)\, .
\]
Thus $f^{k_0}(\omega) = -4\pi^2 T(\omega|k_0|, k_0)$.

We remark, that  in the so called naive scattering theory
$f^{k_0}(\omega)$ is called the scattering amplitude since Born's
ansatz offers also a heuristic way of defining a cross section.
One simply uses the stationary solutions of Schr\"odinger's
equation with the asymptotic behavior (\ref{naive}) to obtain the
cross section from the quantum probability flux through $\Sigma$
generated by the scattered wave: The incoming flux has unit
density and  velocity $v=k_0$. In the outgoing flux  generated by
$f^{k_0}(\omega)\frac{\E^{\I|k_0||x|}}{|x|}$ the number
of particles crossing an area of size $x^2 \D\omega$ about an angle
$\omega$  per unit of time is
\[
|k_0| (|f^{k_0}(\omega)|^2/|x|^2)|x|^2
\D\omega\,.
\]
 Normalizing this with respect the incoming flux
suggests the identification of the cross section with
\begin{equation}\label{fcross}
\sigma_{k_0}^{\rm naive}(\Sigma) :
= \int_\Sigma \D \omega\,|f^{k_0}(\omega)|^2
\end{equation}
in agreement with the above. However, such a heuristic derivation
of the formula (\ref{fcross}) for the  cross section, based solely
on the stationary picture of a one particle plane wave function,
is unconvincing \cite{CohenTanudjiHaag}.

\section{The empirical cross section}
\begin{figure}[ht]
\begin{center}
 \hspace{2cm}\epsfxsize=10cm \epsffile{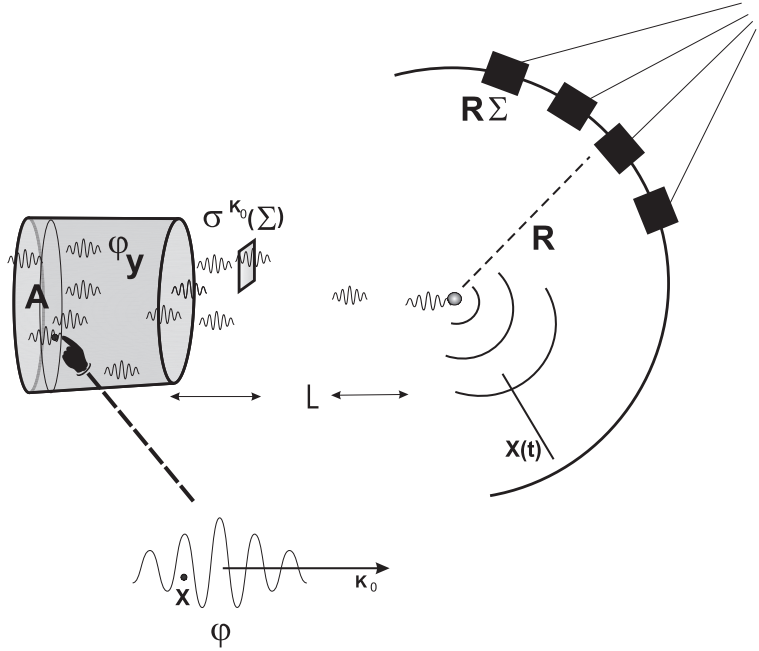}
 \end{center}
\caption{\footnotesize A beam of particles is created in a source
far away (distance $L$) from the scattering center. The
particles' waves are all independent from each other. The
detectors are a distance $R$ away from the scattering center. In
the simplest such models, the wave functions are randomly
distributed over the area $A$ of the beam. The particles arrive
independently at random times at random positions at the detector
surfaces. $\sigma(\Sigma)$ is the cross section, an area which
when put in the incident beam is passed by an equal number of
particles which per unit of time cross the detector surface
defined by the solid angle $\Sigma$. The  random Bohmian position
of the particle within the support of the wave is also depicted as
well as its  straight Bohmian  path $X(t)$ far away from the
scattering center. \label{scattering}}
\end{figure}

Consider a scattering experiment of the most naive kind where one
particle is scattered by a potential. In  Figure~\ref{scattering} we
depict a model for such a scattering experiment, where a beam of
identical independent particles (defining the ensemble) is shot on
a target potential.

 The
scattering cross section for a potential scattering experiment is
measured by the detection rate
 of particles per solid
angle $\Sigma$  divided by the flux $|j|$ of the incoming beam.
 $\Delta T$ is the total time of duration of the
measurement. With $N(\Delta T, R \Sigma)$ denoting the {\it
random} number of particles crossing the surface of the detector
located within the solid angle $\Sigma$, the empirical
distribution is

\begin{equation}\label{em}
  \rho(\Delta T,\Sigma):=\frac{N(\Delta T, R \Sigma) }
  {\Delta T \, |j|}\,.
\end{equation}

The empirical distribution is a {\bf \it random variable} on the
space of ''initial conditions'': initial position of the wave
packet within the beam, time of creation of wave packet, and also
of the {\it quantum randomness}, encoded in the $|\varphi|^2$
randomness. It also depends (in fact very much so) on the
parameters capturing the physical situation, like the distances
$L,R$ and the area $A$ of the beam. The difficult part of this
random variable is the dependence on  the {\it quantum
randomness}, which, as we shall show, becomes simple in the  limit
of large distances. We wish to stress, that the classical
randomness (position of the wave function within the beam, time of
creation of the wave function) which arises from the preparation
of the beam and which in classical scattering theory is all the
randomness there is, adds by virtue of the typical dimensions of the experiment very little to the scattering
probabilities in quantum scattering theory (see \cite{DGTZphysica}
for more on that).

The goal of   scattering theory is to predict the theoretical
value of (\ref{em}). The value predicted is  (\ref{thpre}) or if
one so wishes (\ref{fcross}).

What needs to be shown is thus that, in the sense of the law of
large numbers,
\begin{equation}\label{theorem}
"\lim_{t\to\pm\infty}" \lim_{\Delta T \to \infty} \rho(\Delta
T,\Sigma)= \sigma_{k_0}(\Sigma)\,,
\end{equation}
where the law of large numbers (contained in $\lim_{\Delta T \to \infty}$)  will have to be formulated with the
measure on the space of the initial conditions. The
 ``$\lim_{t\to\pm\infty}$''  refers to large distance limits and limits
 which make the expression beam-model independent:
 \begin{equation}\label{lim}
 "\lim_{t\to\pm\infty}"= \lim\limits_{|\widehat\varphi(k)|^2\to\delta(k-\bko)}
  \lim\limits_{L\to\infty}
  \lim\limits_{|A|\to\infty}\lim\limits_{R\to\infty}
  \end{equation}

In particular the limit  $\lim\limits_{R\to\infty}$ is taken to
obtain the ``local plane wave'' structure (see (\ref{Doll})) of
the scattered wave, which allows for a particular simple
expression for the crossing probability of a particle through the
detector surface. For more explanations of the limits see
\cite{DGTZphysica,Moser}.

\section{The heuristics of quantum randomness}

The random number $N(\Delta T, R \Sigma)$ defining (\ref{em}) is the random sum of ``independent'' single particle contributions, i.e.\
 it depends on the ``trivial'' randomness arising from the
beam, which is simply ensuring the independence of the single detections in the ensemble
 for the law of large numbers to hold. Most importantly, however, it depends on the quantum randomness
inherent in a single event. We shall from now on focus on the scattering of one single particle and
forget the beam. One particle is send towards the scattering center.
The question we must then answer is: Which detector clicks? We must answer this question for the real situation
where the detectors
are a finite distance away from the scattering center. The answer might be complicated but it is that answer
of which one can then take the mathematical limit of infinite distances to obtain a simpler looking formula.

\begin{figure}[ht]
\hspace{2cm}\epsfxsize=8cm \epsffile{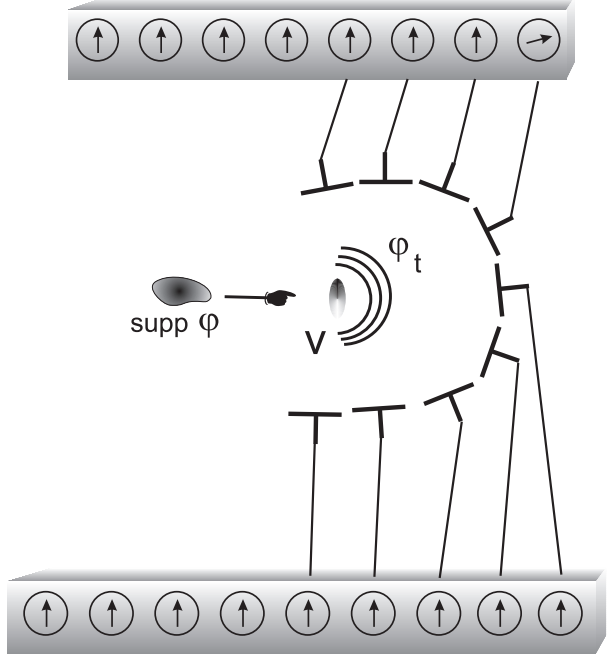}
\caption{
\footnotesize Which detector clicks
 when? The detection time $T_{\rm e}$  and position $X_{\rm e}= X(T_{\rm e})$  are {\it
 random} exit time and exit position.\label{figure2}}
 \end{figure}

Once this question is clear one immediately sees that this question is coarse grained, it  already
ignores that the time at which the particle is registered is random too. The fundamental question  is:
{\em Which detector clicks when?} In other words: What is the distribution for the first exit time and exit position of the
particle from the region defined by the detector surfaces (see Figure~\ref{figure2}).

\begin{equation}\label{exitP}
\mathbb{P}^\varphi\big( X(T_{\rm e})\in \D \Sigma, T_{\rm e} \in \D t
 \big)=\,\,{\bf  ?}\,\,.
 \end{equation}

\begin{figure}[ht]\label{figure3}
\hspace{4.0cm}\epsfxsize=7cm \epsffile{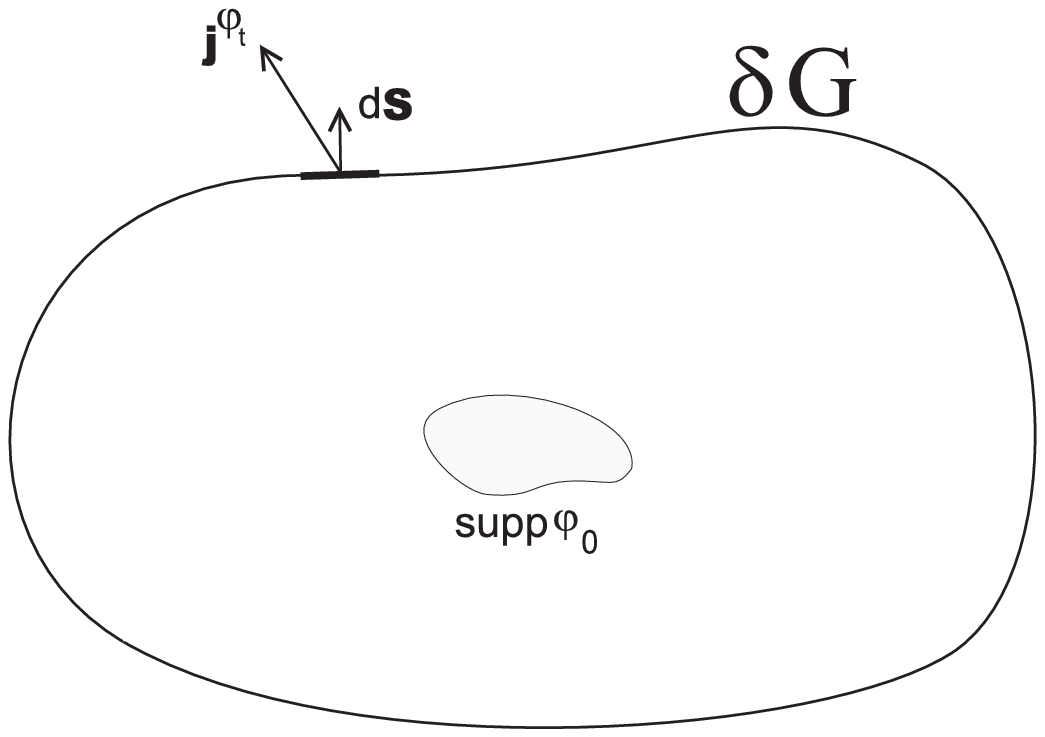}
\caption{\footnotesize Escape of the particle from the region $G$. When the boundary $\partial G$ is far from
the initial support of the wave function, the exit statistics are approximated by the flux through the surface. }
\end{figure}

Heuristically it is clear that the probability is given by the
quantum flux through the surfaces. The quantum flux is
$${ j}^{\varphi_t} = {\rm Im}\,\varphi_t^* \nabla \varphi_t, $$ and appears in an identity---the
so called quantum flux equation---that holds for any $\varphi_t$
being a solution of Schr\"odingers equation:
\begin{equation}\label{quflux}
 \frac{\partial |\varphi_t|^2}{\partial t} + {\rm div   }  \, {
j}^{\varphi_t} = 0\,.
\end{equation}
Consider like in Figure~\ref{figure3} the escape of a  particle initially localized in $G$
through a section $\D S $ of the boundary $\partial G$ (we can but need not think of a freely evolving wave).
If the surface is   far away from the scattering region, it is   very suggestive that the
probability should be given by the flux integrated against the surface
\begin{equation}\label{exitapprox}
\mathbb{P}^\varphi\big( X(T_{\rm e})\in\D S, T_{\rm e} \in \D t
\big)\approx \lim_{|R|\to
\infty}{{j}^{\varphi_t}({R},t) \cdot \D{ S}}\D t\,.
\end{equation}
Based on  this heuristic connection the   flux across
surfaces theorem, which we formulate here in a lax manner, becomes
a basic assertion in the foundations of scattering theory
\cite{AmreinPearson,FASBDT,PT,DP,DGTZphysica}. By integrating the flux against the surface integral
over all times, we ignore the time at which the particle crosses
the surface and we focus merely on the direction in which the
particle moves:\\[15mm]

\noindent {\bf Theorem ``FAST''}: Let $\varphi$ be a (smooth) scattering
state, then

\begin{eqnarray}\label{fast} \lim_{R \to
\infty} \int_0^\infty \D t \int_{R\Sigma} {j}^{\varphi_t} \cdot
\D{S} =\lim_{R \to \infty} \int_0^\infty \D t \int_{R\Sigma}
|{ j}^{\varphi_t} \cdot \D { S}| \nonumber\\ \\\nonumber =\int_{C_\Sigma}
\D k\, \,|\widehat{W_+^*\varphi}({k})|^2\,.
\end{eqnarray}

The heuristics of the FAST is easy to grasp. If we think of a freely evolving wave packet then its long time
asymptotic (which goes hand in hand with a long  distance asymptotic) is (recall $\frac{\hbar}{m}=1$)
\begin{equation}\label{Doll}
e^{-itH_0}\varphi({x})\approx
\frac{\E^{\I\frac{x^2}{2t}}}{t^{\frac{3}{2}}}\,\widehat \varphi\left(\frac{{ x}}{t}\right)\,.
\end{equation}
We call this approximation the local plane wave approximation. It corresponds to a radial outward pointing flux.
For scattering states $\varphi$ of (short range) potential scattering there exists   a
state $\varphi_{\rm out}$ moving freely, so that
\[\lim_{t\to\infty}\|\E^{-\I Ht}\varphi - \E^{-\I H_0t}\varphi_{\rm out}\|=0\,
\]
which leads to the wave operator
\[W_+ := \mbox{s-}\lim_{t\to\infty}\E^{\I Ht}\E^{-\I H_0t}\]
with
\[W_+^* \varphi=\varphi_{\rm out}\,.\]
Combining this with (\ref{Doll}) and computing the flux for this
approximation yields that the left hand side of (\ref{fast})
equals the right hand side of (\ref{fast}). We note that the first
equality  in (\ref{fast}) asserts, that the flux is outgoing, a
condition of vital importance for its interpretation as crossing
probability. We shall discuss its importance below. We remark,
that the further treatment of the right hand side of (\ref{fast})
is more or less standard and becomes upon averaging over the beam
statistics essentially (\ref{thpre}) \cite{Amrein, DGTZphysica, Moser}. That
is, given the FAST, the connection with the $S$-matrix formalism is
standard. The cross
section is justified in the sense of the law of large numbers, once
(\ref{exitapprox}) is accepted.

\section{Bohmian~Mechanics~and~the~justification~of~(\ref{exitapprox})}

The foregoing discussion is necessarily unprecise since the
fundamental objects {\it exit time} and {\em exit position} remain
undefined: There is no time dependent position of the particle in
quantum theory defining these random variables. In Bohmian
mechanics, e.g.\ \cite{BohmIntro}, when the wave function is
$\varphi_t$, there is a particle, and  the particle moves along a
trajectory ${{X}}(t)$ determined by the differential equation
\begin{equation}\label{Bohm1}
\frac{\D}{\D t} {{X}}(t) = v^{\varphi_t}({{X}}(t)) := {\rm Im}\,
\frac{{\bf \nabla} \varphi_t}{\varphi_t}({{X}}(t)),
\end{equation}
Its position at time $t$ is randomly distributed according to the
probability measure $\mathbb{P}^{\varphi_t}$ having density
$\rho_t=|\varphi_t|^2$, see \cite{DGZ}.

The continuity equation  for the probability transport along the
vector field $ v^{\varphi_t}(x,t)$  becomes for the particular
choice $\rho_t= |\varphi_t|^2$ the quantum flux equation
(\ref{quflux}), which establishes that $|\varphi_t|^2$ is an
equivariant density.

Hence the trajectories ${{X}}(t,{{X}}_0)$ are random trajectories,
where the randomness comes from the
$\mathbb{P}^{\varphi}$-distributed random initial position
${{X}}_0$, with $\varphi$ being the ``initial'' wave function.
Having this, the escape time and position problem (\ref{exitP}) is
readily answered. Define $T_{\rm e}= \inf\{t|\, X(t)\in G^c\}$ and
put $X_{\rm e}=X(T_{\rm e})$, then both variables are random
variables on the space of initial positions of the particle and
$\mathbb{P}^{\varphi}(\{X|\, T_{\rm e}(X)\in \D t , \,\,X(T_{\rm
e}(X),X)\in \D S\})$ is clearly the exit distribution we are
looking for. Note also, that we may now specify rigorously the
probability space on which the empirical distribution $(\ref{em})$
is naturally defined, and we furthermore have the measure, with
which the law of large numbers (\ref{theorem}) can be proven.

\begin{figure}[t]
\begin{center}
 \hspace{2cm}\epsfxsize=10.5cm \epsffile{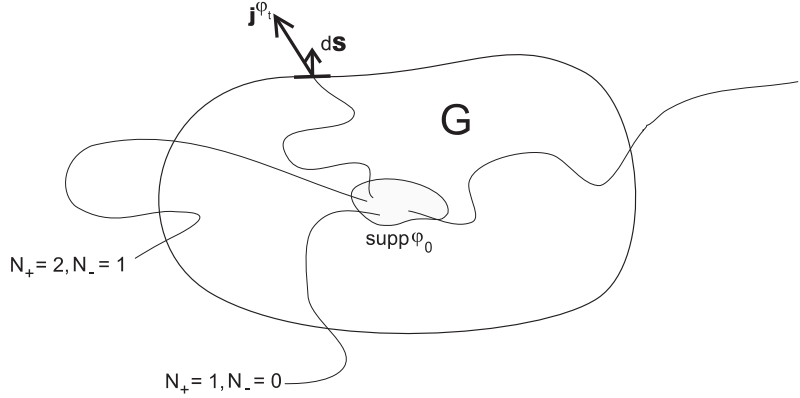}
 \end{center}
\caption{\footnotesize Signed number of crossings of possible
trajectories through the boundary of the region $G$.
\label{figure4}}
\end{figure}

We explain now the connection of this exit probability with the
flux. Consider some possible exit scenarios of the particle as in
Figure~\ref{figure4}.
We introduce the random variables {\em number of crossings}
\[
N(\D S,\D t) :=  N_+(\D S,\D t)
 + N_-(\D S,\D t)\,.
 \]
and  {\em number of signed crossings}
\[
N_s(\D S,\D t) := N_+(\D S,\D t) -
N_-(\D S,\D t)\,,
\]
where $N_\pm(\D S,\D t)$ are the number of outward resp.\ inward crossings.
 Their expectations are readily computed
in the usual statistical mechanics manner: For a crossing of $\D
S$ in the time interval $(t,t+\D t)$ to occur, the particle has to
be in a cylinder (Boltzmann collision cylinder) of size
$|v^{\varphi_t}  \cdot \D{ S}\,\D t| $ at time $t$. Thus
\[
\mathbb{E}^\varphi(N(\D S,{\rm d}t)) = |\varphi_t|^2
\,|v^{\varphi_t} \cdot \D{ S}|\,\D t = |{ j}^{\varphi_t}
\cdot \D{ S}|\,\D t
\]
and
\begin{equation} \label{Eflux}
 \mathbb{E}^\varphi(N_s(\D S,\D t)) =
 {j}^{\varphi_t} \cdot \D{S} \,\D t\,.
 \end{equation}
Under the condition that the flux is positive for all times
through the boundary of $G$ (a condition which needs to be
proven, and which is asserted in the first equality of
(\ref{fast}))
  every trajectory crosses
the boundary of $G$ at most once. Hence
\begin{eqnarray}\nonumber
 \mathbb{E}^\varphi(N(\D S,{\rm
d}t) )= \mathbb{E}^\varphi(N_s(\D S,\D t))
  = 0\cdot\mathbb{P}^\varphi(T_{\rm e}\notin \D t {\mbox{\rm \ or }} {X}_e\notin \D S) +\\
\nonumber 1 \cdot \mathbb{P}^\varphi({X}_e\in \D S {\mbox{\rm \ and }} T_{\rm e}\in \D t)\,.
\end{eqnarray}
In that particular situation the exit
probability is thus
\begin{equation}\label{pf}
\mathbb{P}^\varphi({X}_e\in {\rm
d}S {\mbox{\rm \ and }} T_{\rm e}\in \D t) =
 { j}^{\varphi_t}  \cdot \D{ S} \,\D t\,.
\end{equation}

This and (\ref{fast}) are at the basis of quantum mechanical
scattering theory for single particle potential scattering.

\section{Multi-time distributions for many particles}

We extend the foregoing to the case of many particle scattering.
We shall discuss some of the main steps, which need to be filled
with rigorous mathematics in future works. For simplicity we
consider the free case where the particles are guided by an
entangled wave function, but they do not interact via a potential
term in the Hamiltonian with each other.
However, the following naturally generalizes to interacting particles
by replacing the wave function $\varphi$ by its free outgoing asymptote
$\varphi_{\rm out} = W^*_+\,\varphi$.
While Bohmian mechanics
naturally extends to many particles (see (\ref{multibohm}) below),
one sees immediately that our  task of getting our hands on the
exit statistics for many particles  is nevertheless nontrivial,
since every particle has its own exit time and position. I.e.\ we
need to handle
\begin{equation}\label{multiex}
\mathbb{P}^{\varphi}\big(T_{\rm e}^{(1)}\hspace{-1pt}\in {\rm
d}t^{(1)},{X}^{(1)}\hspace{-1pt}(T_{\rm e}^{(1)})\hspace{-1pt}\in \D
S^{(1)},\ldots,T_{\rm e}^{(n)}\hspace{-1pt}\in \D t^{(n)}, {X}^{(n)}\hspace{-1pt}(T_{\rm e}^{(n)})\hspace{-1pt}\in
\D  S^{(n)}\big)\,.
\end{equation}

To apply the statistical mechanics argument which we used in the
last section to compute the crossing probability  the multi-time
position distribution is needed
\begin{eqnarray}\label{multiti}\lefteqn{\hspace{-2cm}
 \mathbb{P}^{\varphi}\big( X^{(1)}(t^{(1)})\in \D  {x}^{(1)}
,\ldots,X^{(n)}(t^{(n)})\in \D
{x}^{(n)}\big)}\\&   &=\rho({x}^{(1)},t^{(1)},\ldots,{x}^{(n)},t^{(n)})\,\D {x}^{(1)} \ldots \D  {x}^{(n)}\,,\nonumber
\end{eqnarray}
which in general will not be a simple functional of the wave
function. We will show   that in the scattering regime,
when the wave approaches the local plane wave structure, this multi-time
position distribution can be computed and the exit statistics are in
fact given by a particular multi-time flux form.  To our best knowledge, this
observation is new. The single-time multi particle flux has been
used in \cite{CNS} to compute exist statistics, necessarily ignoring particle correlations.

For ease of notation we consider  two particles with positions
$X,Y$ and wave function $\varphi(x,y,t).$ The Bohmian law of
motion is

\begin{eqnarray}\label{multibohm}
&&\dot{X}(t)= v^x_t(X(t),Y(t))=\Im \frac{\nabla_x
\varphi(x,y,t)}{\varphi(x,y,t)} \Big|_{x=X(t),\,y=Y(t)}\\[2mm]
&&\dot{Y}(t)= v^y_t(X(t),Y(t))=\Im \frac{\nabla_y
\varphi(x,y,t)}{\varphi(x,y,t)} \Big|_{x=X(t),\,y=Y(t)}\\[2mm]
&&\I  \partial_t  \varphi(x,y,t)=-
 {\textstyle\frac{1}{2}}\big(\Delta_x+\Delta_y\big)\varphi(x,y,t).
\end{eqnarray}

With $H=H_x+H_y=-\frac{1}{2}\big(\Delta_x+\Delta_y\big)$  we can
easily produce a two times wave function by the appropriate action of
the single particle Hamiltonians through
\begin{equation}\label{multiwave}
\varphi(x,t,y,s):= \big( \E^{-\I H_x t} \E^{-\I H_y s} \varphi_0 \big) (x,y)\,,
\end{equation}
which reduces to the usual single-time wave function for $t=s$, because the Hamiltonians $H_x$ and $H_y$ commute.
Hence one could as well include single particle potentials into  $H_x$ and $H_y$.
While the definition of $\varphi(x,t,y,s)$ seems very natural at first sight,
note that the physical meaning of $|\varphi(x,t,y,s)|^2$ is not at all obvious.
To get our hands on this question, let
\[
\Phi_t(x,y)=\big(\Phi^x_t(x,y),\Phi^y_t(x,y)\big)=
\big(X(t,x,y),Y(t,x,y)\big)
\]
be the Bohmian flow along the vector field
given by (\ref{multibohm}) transporting the initial values $x,y$
along the Bohmian trajectories  to values at time $t$  and let
\[
\Phi_{t,s}(x,y)=\big(\Phi^x_t(x,y),\Phi^y_s(x,y)\big)=
\big(X(t,x,y),Y(s,x,y)\big)
\]
 be the two  times Bohmian  flow. Observe
that
\begin{eqnarray}\label{abl}
\partial_t \Phi_{t,s}(x,y)&=&\big(\partial_t
\Phi^x_t(x,y),0\big)=\Big(v^x_t\big(\Phi_t(x,y)\big),0\Big)\\
\partial_s \Phi_{t,s}(x,y)&=&\big(0, \partial_s
\Phi^y_s(x,y)\big)=\Big(0,v^y_s\big(\Phi_s(x,y)\big) \Big)\,  .
\end{eqnarray}

>From the definition of the multi-time wave function
(\ref{multiwave}) it follows in the same  way as in the
single-time case that
\begin{eqnarray}\label{multifluxx}
\partial_t|\varphi(x,t,y,s)|^2 &=& -\nabla_x \cdot {\rm Im} \big(\varphi(x,t,y,s)^*\nabla_x\varphi(x,t,y,s)\big) \nonumber\\
\partial_s|\varphi(x,t,y,s)|^2 &=& -\nabla_y\cdot {\rm Im }\big(\varphi(x,t,y,s)^*\nabla_y\varphi(x,t,y,s)\big)\,,
\end{eqnarray}
which leads us to define a multi-time velocity field:
\begin{equation}\label{multifluxyIm}
v^x_{t,s}(x,y) = \Im \frac {\nabla_x
\varphi(x,t,y,s)}{\varphi(x,t,y,s)}
\end{equation}
if $\varphi(x,t,y,s)\ne 0$ and $v^x_{t,s}(x,y)=0$ if
$\varphi(x,t,y,s)=0$ and analogously for $ v^y_{t,s}(x,y)$.

We show now, that under certain conditions there exists a  two times continuity equation for a two times density
$\rho(x,t,y,s)$. We start with the definition,
setting $\rho(x,0,y,0)=\rho(x,y)$,
\begin{eqnarray}\label{twotimedef}
\mathbb{E}^\varphi\Big(f\big(X(t),Y(s)\big)\Big)&=&\int\D x\D y\,
f\big(\Phi_{t,s}(x,y)\big)\rho(x,y)\nonumber\\&=:& \int\D x\D y\,
f(x,y)\rho(x,t,y,s)\,,
\end{eqnarray}
where $f$ varies in  a suitable class of test functions.
Next differentiate the equation with respect to $t$ respectively
$s$. This yields in the second equality
\begin{eqnarray}\label{rechnung}\lefteqn{\hspace{-10pt}
\partial_t\int \D x\D y\,f\big(\Phi_{t,s}(x,y)\big)\rho(x,y)}\nonumber\\&=& \int\D x\D y\,\nabla_{(1)}
f\big(\Phi_{t,s}(x,y)\big)\cdot v^x_t\big( \Phi_t(x,y)\big) \rho(x,y)\nonumber \\&=&\int\D x\D y\, f(x,y)\partial_t\rho(x,t,y,s)\,,
\end{eqnarray}
and similarly for differentiation with respect to $s$. Here
$\nabla_{(1)}$ denotes the gradient with respect to the first
argument. If the following ``multi-time independence'' condition
\begin{eqnarray}\label{cond}
 v^x_t\big( \Phi_t(x,y)\big)&=& v^x_{t,s}
\big(\Phi^x_t(x,y), \Phi^y_s(x,y)\big)\nonumber\\[-5pt]\\[-5pt]\nonumber v^y_t\big(
\Phi_t(x,y)\big)&=& v^y_{t,s} \big(\Phi^x_t(x,y),
\Phi^y_s(x,y)\big)
\end{eqnarray}
is satisfied, we can replace $v^{x}_t\big(
\Phi_t(x,y)\big),v^y_t\big( \Phi_t(x,y)\big)$ in (\ref{rechnung})
by
\[
v^{x}_{t,s} \big(\Phi^x_t(x,y), \Phi^y_s(x,y)\big)\,.
\]
 Using definition (\ref{twotimedef}) followed by partial integration yields for the
 second integral in (\ref{rechnung})
\begin{eqnarray*} \lefteqn{
\int\D x\D y \,\nabla_{(1)}
f\big(\Phi_{t,s}(x,y)\big)\cdot v^x_t\big( \Phi_t(x,y)\big) \rho(x,y)   }\\&=&
\int\D x\D y \,\nabla_{(1)}
f\big(\Phi_{t,s}(x,y)\big)\cdot v^x_{t,s}
\big(\Phi_{t,s}(x,y) \big) \rho(x,y) \\&\stackrel{(\ref{rechnung})}{=}&
\int\D x\D y\,\nabla_{(1)}
f\big( x,y\big)\cdot v^x_{t,s}
\big(x,y \big) \rho(x,t,y,s)
  \\&=&
-\int\D x\D y\, f(x,y)\,\nabla_x\cdot \left( v^x_{t,s}( x,y) \rho(x,t,y,s)\right)\,.
\end{eqnarray*}
>From this and (\ref{rechnung}) we may conclude, repeating the same
for the $s$-differen\-tia\-tion,  the two times continuity equation
\begin{eqnarray}\label{twotimecont}
\partial_t\rho(x,t,y,s)&=&-\nabla_x \cdot\left(
v^x_{t,s}(x,y)\rho(x,t,y,s)\right)\nonumber\\
\partial_s\rho(x,t,y,s)&=&-\nabla_y \cdot\left(
v^y_{t,s}(x,y)\rho(x,t,y,s)\right)\,.
\end{eqnarray}
Comparing this with (\ref{multifluxx}) we see that
$\rho(x,t,y,s)=|\varphi(x,t,y,s)|^2$ is equivariant. All this
depends crucially on the ``multi-time independence'' condition
(\ref{cond}). It is easy to see that the condition is satisfied if
the wave function is a product wave function. But that is
uninteresting. The condition can be expected to be also
approximately satisfied when the wave function attains the local
plane wave structure
\begin{equation}\label{Dolltwo}
\varphi(x,t,y,s)\approx \frac{\E^{\I\frac{x^2}{2t}}}{t^{\frac{3}{2}}}\,
\frac{\E^{\I\frac{y^2}{2s}}}{s^{\frac{3}{2}}}\,
 \widehat\varphi\big(\frac{x}{t},\frac{y}{s}\big)
\end{equation}
of an outgoing scattering state at large times (see next section).
In this case the trajectories are approximately straight lines and
the velocity of particle $X$ does not change if particle $Y$ is
moved along its straight path and vice versa. We remark that the
local plane wave structure is preserved under multi-time evolution
(as it is preserved under single time evolution). Thus   in the scattering regime condition (\ref{cond}) holds
true and we conclude that in this regime the two-times wave
function (\ref{multiwave}) yields the two-times joint distribution
$\rho(x,t,y,s)=|\varphi(x,t,y,s)|^2$ for the positions of the two
particles. Hence, approximately,  we have that
\[
\mathbb{P}^\varphi \big( X(t) \in \Lambda_1 \, \mbox{and}\,Y(s) \in \Lambda_2 \big) \approx
\int_{\Lambda_1} \D x \int_{\Lambda_2}\D y \,|\varphi(x,t,y,s)|^2\,.
\]

Moreover we have in that regime single crossings only. We can thus
compute the exit statistics in the scattering regime like before
(the Boltzmann collision cylinder argument) but now using the two
times density $|\varphi(x,t,y,s)|^2$ and the approximate straight
path veloci\-ties
\begin{equation}
v^x_{t,s}(x,y) \approx \frac{x}{t}\,\,\,,  v^y_{t,s}(x,y) \approx
\frac{y}{s}
\end{equation}
 This way
one obtains
\begin{eqnarray}\lefteqn{\hspace{-1cm}\label{jprob}
\mathbb{P}^\varphi(T^x_{\rm e}\in\D t, T^y_{\rm e}\in\D s, X(T^x_{\rm e})\in
\D S^x , Y(T^y_{\rm e})\in\D S^y)}
\\&\approx& |\hat{\varphi}\big(\frac{x}{t},\frac{y}{s}\big)|^2 \,
\big(\frac{x}{t}\cdot \D S^x\big)\,\big(\frac{y}{s} \cdot \D S^y\big)\, \D t\,\D s\nonumber
\\&\approx:&
j^{\rm sp}(x,t,y,s)\cdot (\D S^x \otimes \D S^y)  \, {\rm
d}t\,\D s\,,\nonumber
\end{eqnarray}
where the two-times ``straight paths''    flux form
$j^{\rm sp}(x,t,y,s)$ is the straight path approximation to the
multi-time flux form
\begin{equation}\label{multitimeflux}
j(x,t,y,s):=|\varphi(x,t,y,s)|^2\,\, v^x_{t,s}(x,y) \otimes
v^y_{t,s}(x,y)   \,.
\end{equation}

It is remarkable and relevant for its meaning in the foundations
of scattering theory that this {\em unmeasured} Bohmian joint
probability is in this particular situation the same as the
measured probability, which is in general not true for joint
probabilities \cite{Berndl}. Measurements lead---in the language
of orthodox quantum theory---to a collapse of the wave function,
which in the local plane wave approximation however does
 not have any effect on the trajectory of the other
particles. In the two particles case the collapse (due to the
detection of one particle) picks out simply the rightly correlated
pair, which in fact can be EPR correlated pairs.

 The $N$-particle multi-time flux (\ref{multitimeflux}) as well as the
$N$-particle single time flux have taken alone no significance for
the description of scattering (in contrast to the one particle
situation), while the crossing probabilities (\ref{jprob}) of
course do. We shall in the next section compute the value of the
right hand side of (\ref{jprob}), which is the usual scattering
into cones (in momentum space) formula.

\section{The exit statistics theorem for $N$ particles}

We abbreviate the joint exit time-exit position distribution for
$N$ particles through a sphere of radius $R$ as
\begin{eqnarray*}\lefteqn{
\mathbb{P}^\varphi( \D t_1\ldots\D t_N\,\D S_1\ldots\D S_N):=}\\&&
\mathbb{P}^\varphi(X_{1}(T_{1\E})\in\D S_1, T_{1\E}\in \D
t_1,\ldots, X_{N}( T_{N\E})\in\D S_1, T_{N\E}\in \D t_N)\,,
\end{eqnarray*}
where we recall that $T_{n\E}$ is the first exit time of the $n$th particle through  the sphere and
$\D S_n$   an infinitesimal surface element on this sphere.
Neglecting the possibility of clustering,
the generalization of the flux-across surfaces theorem of potential scattering then becomes the following conjecture.
\bigskip

\noindent {\bf Exit Statistics Theorem}: {\em Let $\varphi$ be a
(smooth) scattering state of an   $N$-body Hamiltonian $H$ at time
$t=0$, then  for any $-\infty<T<\infty$
\begin{eqnarray}\lefteqn{
\lim_{R\to\infty}\int_T^\infty \cdots\int_T^\infty
\int_{R\Sigma_1}\cdots \int_{R\Sigma_N}
\, \mathbb{P}^\varphi( \D t_1\ldots\D t_N\,\D S_1\ldots\D S_N) =}\nonumber\\
&=& \lim_{R\to\infty}\int_T^\infty\hspace{-4pt}\D t_1\cdots
\int_T^\infty\hspace{-4pt}\D t_N \int_{R\Sigma_1}
\hspace{-4pt}\cdots  \int_{R\Sigma_N}\hspace{-1pt} j^{\varphi_{\rm
out,sp}}(x_1,\ldots,x_N,t_1,\ldots,t_N)\nonumber\\&&
\hspace{7,5cm}\cdot
 (\D S_1\otimes..\otimes \D S_N)\nonumber\\
  \nonumber\\ \label{NFAST}
&=& \int_{C_{\Sigma_1}} \D k^3_1\cdots\int_{C_{\Sigma_N}} \D
k^3_N\,|\widehat \varphi_{\rm out}(k_1,\ldots,k_N)|^2\,.
\end{eqnarray}}\bigskip

Recall that $\varphi_{\rm out} = W_+^* \varphi$ and that
\[
\varphi_{\rm out}(t_1,\ldots,t_N)=\E^{\I\Delta_{x_1} t_1}\cdots
\E^{\I\Delta_{x_N} t_N}\varphi_{\rm out}
\]
evolves according to the free multi-time evolution.

The theorem provides a precise connection between the joint
distribution of the measured exit positions of $N$ scattered
particles (the first expression in (\ref{NFAST})) and the
empirical formula for this quantity in terms of the Fourier
transform of the outgoing wave (the last expression in
(\ref{NFAST})). A rigorous proof of this connection seems to
involve necessarily a multi-time formulation of the quantum
mechanics in the scattering regime in the sense of the
intermediate expression in (\ref{NFAST}). Notice that the first
equality in (\ref{NFAST}) is, as discussed in the previous
section, the highly nontrivial part to prove. More precisely, one
needs to establish (\ref{jprob}) rigorously and with error
estimates which are integrable in the sense of (\ref{NFAST}). The
second equality in (\ref{NFAST}) is an easy computation, with
which we shall conclude the paper. We shall first remind the
reader of the local plane wave structure which approximates the
scattering state and which is presumably crucial for the proof of
the theorem.

Since $|\widehat\varphi_{\rm out}(k)|$ is invariant under the free
time-evolution  we can choose without loss of generality $T\geq
1$. To shorten notation lets introduce the configuration variables
$\overline x=(x_1,\ldots, x_N)$ and $\overline
t=(t_1,\ldots,t_N)$. Then
\begin{eqnarray*}
\varphi_{\rm out}(\overline x,\overline t)&=&\left(\E^{\I\Delta_{x_1} t_1}\cdots \E^{\I\Delta_{x_N} t_N}\right)\varphi_{\rm out}(\overline x)\\
&=& \hspace{-3pt}\int_{\RRR^3}\hspace{-1pt} \D y_1\cdots\int_{\RRR^3}\hspace{-1pt}\D y_N
\frac{\E^{\I\frac{|x_1-y_1|^2}{2t_1}}}{(2\pi \I t_1)^\frac{3}{2}}
\cdots \frac{\E^{\I\frac{|x_N-y_N|^2}{2t_N}}}{(2\pi \I
t_N)^\frac{3}{2}}\,\varphi_{\rm out}(y_1,\ldots,y_N) \,,
\end{eqnarray*}
where here and in the following $\varphi_{\rm out}$ without a
time-argument means always $\varphi_{\rm out}(\overline t=0)$.
Expanding every factor in the integrand as
\[
 \E^{\I \frac{|x_n-y_n|^2}{2t_n}} = \E^{\I \frac{|x_n|^2}{2t_n}} \,\E^{-\I\frac{x_n\cdot y_n}{t_n}}
 + \E^{\I \frac{|x_n|^2}{2t_n}} \,\E^{-\I\frac{x_n\cdot y_n}{t_n}}
 \left( \E^{\I \frac{|y_n|^2}{2t_n}} -1\right) \,,
\]
one obtains
\begin{equation}\label{NAsymp}
\varphi_{\rm out}(\overline x,\overline t)=\frac{\E^{\I
\frac{|x_1|^2}{2t_1} }}{( i t_1)^\frac{3}{2}}\cdots\frac{\E^{\I
\frac{|x_N|^2}{2t_N} }}{(i t_N)^\frac{3}{2}  } \,\widehat
\varphi_{\rm out}\left( \frac{x_1}{t_1},\ldots,
\frac{x_N}{t_N}\right)+ R(\overline x,\overline t)\,,
\end{equation}
where every term in the sum  $R$ has at least one factor of the form
\[
\left( \E^{\I \frac{|y_n|^2}{2t_n}} -1\right)
\]
 in the integrand. Under appropriate assumptions on $\varphi_{\rm out}$ it is now easy
 to get estimates on the remainder term $R(\overline x,\overline t)$ for large $t_n$ by stationary phase methods.
 For details we refer to \cite{DT}.
 In particular the remainder term does not contribute to the time integrals in (\ref{NFAST}).

 Neglecting $R$ we obtain from (\ref{NAsymp})  for the $n$th component of the velocity
 \begin{equation}\label{vout}
 v^n_{\overline t}(\overline x)= \frac{x_n}{t_n} + \frac{1}{t_n}{\rm Im}\,\frac{\nabla_{n} \widehat \varphi_{\rm out}\left( \frac{x_1}{t_1},\ldots, \frac{x_N}{t_N}\right)
 }{\widehat \varphi_{\rm out}\left( \frac{x_1}{t_1},\ldots,
 \frac{x_N}{t_N}\right)},
 \end{equation}
 of which we only need the first term (the straight path velocity)
 and for the density
 \[
 | \varphi_{\rm out}(\overline x,\overline t)|^2 = \frac{1}{t_1^3\cdots t_N^3} \left|\widehat \varphi_{\rm out}\left( \frac{x_1}{t_1},\ldots, \frac{x_N}{t_N}\right)\right|^2\,.
 \]

 Using $x_n\cdot \D S_n = |x_n| R^2 \D \omega_n = R^3 \D \omega_n $, where $\D \omega$ denotes
 Lebesgue measure on the unit sphere $S^2\subset \RRR^3$,
we now conclude with the computation of the second equality of
(\ref{NFAST}):
 \begin{eqnarray*}\lefteqn{
 \lim_{R\to\infty}\int_T^\infty\hspace{-8pt}\D t_1\cdots \int_T^\infty\hspace{-8pt}\D t_N
\int_{R\Sigma_1} \hspace{-8pt}\cdots \int_{R\Sigma_N}\hspace{-5pt}
j^{\varphi_{\rm out,sp}}(\overline x,\overline t)\cdot
 (\D S_1\otimes\ldots\otimes \D S_N)}\\
&=&
 \lim_{R\to\infty}\int_T^\infty\hspace{-8pt}\D t_1\cdots \int_T^\infty\hspace{-8pt}\D t_N
\int_{R\Sigma_1} \hspace{-10pt}\cdots \int_{R\Sigma_N}
\frac{ \left|\widehat \varphi_{\rm out}\left( \frac{R \omega_1}{t_1},\ldots, \frac{R\omega_N}{t_N}\right)\right|^2  }{t_1^4\cdots t_N^4}   R^{3N}\\ &&\hspace{8,5cm}\D\omega_1\cdots \D\omega_N\\\\
&=&
 \lim_{R\to\infty}\int_0^{\frac{R}{T}}\hspace{-8pt}\D |k_1|\cdots \int_0^{\frac{R}{T}}\hspace{-8pt}\D |k_N|
\int_{R\Sigma_1} \hspace{-8pt}\cdots \int_{R\Sigma_N}
    \left|\widehat \varphi_{\rm out} ( k_1,\ldots,k_N )\right|^2 \\&&\hspace{6.4cm} |k_1|^2\cdots|k_N|^2\,\D\omega_1\cdots \D\omega_N \\\\
    &=&
\int_{C_{\Sigma_1}} \D k^3_1\cdots\int_{C_{\Sigma_N}} \D
k^3_N\,|\widehat \varphi_{\rm out}(k_1,\ldots,k_N)|^2\,.
 \end{eqnarray*}
 In the above computation we substituted $k_n = \frac{x_n}{t_n}$, which, in particular, gives
 $\D t_n = - t_n^2 R^{-1}\D|k_n|$ and $R/t_n = |k_n|$.

\section{Conclusion}

For the first time we formulate the connection between the
joint distribution of the measured exit positions of $N$ scattered particles and the empirical formula
for this quantity in terms of the Fourier transform of the outgoing wave.  While in the case of potential scattering
for a single particle the distribution of the measured exit position can be formulated, at least heuristically,
in terms of the quantum flux, this is no longer true for the joint distribution of $N$ particles. In the case
of $N$ particle scattering even the definition of the relevant distribution is not possible within orthodox
quantum mechanics.  Therefore we use the Bohmian trajectories of the particles to define the distribution
of exit positions and times. The flux-across-surfaces theorem for $N$ particles then connects
this fundamental joint distribution with the empirical formulas of quantum mechanics.
While a completely rigorous proof of the flux-across-surfaces theorem for $N$ particles seems a challenging task,
we sketched a possible argument and showed that a multi-time formulation of the quantum mechanics in the scattering regime should play a
crucial role in this program.

\end{document}